\documentclass[12pt]{iopart_BW}
\usepackage{hyperref}
\usepackage{graphicx}
\usepackage{gensymb}
\usepackage{textcomp}
\usepackage{setspace}
\usepackage[square,comma,sort&compress,numbers]{natbib}
\usepackage{fancyhdr}
\usepackage{setspace}
\usepackage{booktabs}
\usepackage{amsmath}

\begin{document}

\onehalfspacing

\title{Diffuse scattering in Ih ice}

\author{Bj\"orn Wehinger$^1$, Dmitry Chernyshov$^2$,  Michael Krisch$^1$,  Sergey Bulat$^3$, Victor Ezhov$^3$ and Alexe\"i Bosak$^1$,}
\address{$^1$ European Synchrotron Radiation Facility, BP 220 F-38043 Grenoble Cedex 9, France}
\address{$^2$ Swiss-Norwegian Beamlines at European Synchrotron Radiation Facility, Grenoble, France}
\address{$^3$ Petersburg Nuclear Physics Institute, NRC Kurchatov institute, Gatchina, 188300 Saint Petersburg, Russia}

\ead{wehinger@esrf.fr}

\begin{abstract}
Single crystals of ice Ih, extracted from the subglacial Lake Vostok accretion ice layer (3621 m depth) were investigated by means of diffuse x-ray scattering and inelastic x-ray scattering. The diffuse scattering was identified as mainly inelastic and rationalized in the frame of \textit{ab initio} calculations for the ordered ice XI approximant. Together with Monte-Carlo modelling our data allowed reconsidering previously available neutron diffuse scattering data of heavy ice as the sum of thermal diffuse scattering and static disorder contribution. 
\end{abstract}

\submitto{\JPCM}
\maketitle

\pagestyle{fancy}
\setlength{\headheight}{16pt}
\lhead{\textit{Diffuse scattering in Ih ice}}
\rhead{\thepage}
\cfoot{}

\section{Introduction}
The hydrogen-bonded network of water molecules within its crystalline structure results in a complex phase diagram with at least 16 crystalline and at least two amorphous forms of ice showing interesting physical and chemical properties \cite{bartels-rausch_rmp_2012,petrenko_OP_1999}. Ice Ih is the only ice existing in the Earth’s crust, while the metastable ice Ic is suspected to form in the atmosphere \cite{kuhs_pnas_2012,murray_nat_2005}. It has been proposed that snow crystals may start growing at low temperatures from a cubic symmetry with stacking disorder before transforming to the hexagonal structure \cite{kuhs_pnas_2012,kobayashi_TSP_1987}. Stacking faults are indeed frequent also in ice Ih and contribute to diffuse scattering \cite{ogura_HUP_1988}. In the ice-Ih phase the oxygen atoms are arranged in a wurzite structure, and the hydrogen atoms are placed randomly according to Pauling's ice rules \cite{pauling_acs_1935}. The hydrogen disorder, however, affects the positions of oxygen atoms, resulting in a deviation of their crystallographic position by up to several hundredth of an Angstrom \cite{kuhs_CUP_1986}. Anomalous features in the lattice dynamics are responsible for the negative thermal expansion at low temperatures and the softening of transverse acoustic modes, an important concept in pressure- induced amorphisation of ice \cite{straessle_prl_2004}. Phonon dispersion relations, measured by inelastic neutron scattering, are available for D$_2$O \cite{straessle_prl_2004,renker_pla_1969,bennington_pb_1999,fukazawa_jcp_2003}, while for light ice(s) with a large incoherent scattering factor of H only the hydrogen-projected density of states could be measured \cite{klug_prb_1991,schober_pb_1997}. Inelastic x-ray scattering was also applied for the studies of polycrystalline \cite{ruocco_prb_1996} and amorphous \cite{koza_prb_2008} ice, and one dispersion branch was determined for a single crystal \cite{ruocco_prb_1996}.

The other source of lattice dynamics information - thermal diffuse scattering - has not yet fully been explored. The only detailed study of the X-ray diffuse scattering dates from 1949 \cite{owston_ac_1949}, and, due to the limitations of the technique at that time, the experimental data - though amazingly good - are incomplete and the reconstructed patterns are only semi-quantitative.
Neutron diffuse scattering is strongly affected by the hydrogen disorder; this contribution was expected to dominate over the thermal diffuse scattering component at low temperature \cite{li_pmb_1994}. As a consequence, no data on neutron diffuse scattering have been used for the study of the lattice dynamics.

The aim of the present work is to revisit the diffuse scattering in ordinary ice Ih, jointly utilizing thermal diffuse scattering, inelastic scattering and \textit{ab initio} calculations \cite{bosak_zkri_2012}. Different approaches of \textit{ab initio} electronic structure calculations for ice-Ih have been proposed \cite{kuo_jcp_2005} and a lattice dynamics calculation following the proposed approximation is in principle possible. However, such calculations are very demanding. We circumvent this difficulty here to some extent using the proton ordered ice XI phase to model the lattice dynamics. Ice XI is the proton ordered phase of water ice, stable at low temperature. It has an orthorhombic structure with space group Cmc2$_1$ \cite{jackson_pcb_1997}. Such an approximation appears to be reasonably valid in terms of oxygen dynamics, and provides some insight into the hydrogen dynamics and its contribution to the scattering.

\section{Experiment}
Single crystal ice samples were cut from the core extracted from the subglacial Lake Vostok accretion ice layer (depth of 3621 m), where the characteristic grain size is reported to be up to 1 m \cite{jouzel_sci_1999}, significantly extending the definition of "grain". They are probably among the most perfect ice crystals, and diffuse scattering should not be affected by any defects related to mosaicity. The pilot diffuse scattering measurements were performed at BM01A of the Swiss-Norwegian Beam Lines at the European Synchrotron Radiation Facility (ESRF) using a mar345 detector with $\sim$90 s readout time per image frame. As we have found it difficult to protect the crystal against sublimation during the rather long experiment, the follow-up experiment was performed on beamline ID29 at the ESRF which is equipped with a single photon counting pixel detector (PILATUS 6M \cite{kraft_jsr_2009}), allowing for a frame rate of 18 Hz in shutterless mode. In both cases the wavelength was set to 0.700 \AA. A rod-like 2 mm thick crystal was gently shaped by a scalpel blade, glued by a water drop to the glass sample holder, mounted on a rotation stage and held in a cryostream flux (175 $\pm$ 1 K). Prior to the measurement the sample was etched by cold methanol, resulting in a smooth transparent surface. Diffuse scattering patterns were recorded with an increment of 0.1$^\circ$ over an angular range of 360$^\circ$. The orientation matrix refinement was performed using the CrysAlis software package \cite{ox_diff}. For the final reciprocal space reconstructions corrections for polarisation and for solid angle conversion associated with the planar projection were applied. All the test samples were found to be good quality single crystals, showing narrow rocking curves. The absence of diffuse streaks indicates that the crystals are largely stacking-fault free. 
The inelastic x-ray scattering (IXS) experiment was performed on beamline ID28 at the ESRF. The instrument was operated at 17794 eV, providing an overall energy resolution of 3.0 meV full-width-half-maximum (FWHM). Direction and size of the momentum transfer were selected by an appropriate choice of the scattering angle and the crystal orientation in the horizontal scattering plane. The momentum resolution was set to $\sim$0.25 nm$^{-1}$ $\times$ 0.75 nm$^{-1}$ in the horizontal and vertical plane, respectively. Further details of the experimental setup and data treatment procedures can be found elsewhere \cite{kirsch_Springer_2007}. Constant-Q scans were performed with an exposure time $\sim$20 s per energy step.

\section{Calculations}
\textit{Ab initio} lattice dynamics calculations of H$_2$O ice XI were performed using density functional perturbation theory \cite{gonze_prb_1997_2} as implemented in the CASTEP package \cite{refson_prb_2006,clark_zkri_2005}. The general gradient approximation within the plane-wave pseudo-potential formalism was employed using norm conserving pseudo-potentials from the Bennett \& Rappe Pseudo-potential Library \cite{rappe_pseudopotentials}. The geometry was optimized on a 7$\times$7$\times$4 Monkhorst-Pack grid with a plane wave cut-off energy of 820 eV, ensuring the convergence of internal forces to $<$10$^{-3}$ eV$/$\AA. The relaxed lattice parameters for the light ice XI are a = 4.396 \AA, b = 7.654 \AA  \hspace{1pt} and c = 7.188 \AA, which corresponds to a $\sim$6 \% smaller volume than experimentally determined for D$_2$O ice XI \cite{howe_jcp_1989}. Indeed, replacing hydrogen by deuterium causes a volume increase due to anomalous nuclear quantum effects, which are well described within density functional theory \cite{pamuk_prl_2012}. The lattice dynamics calculation of D$_2$O was performed using the same pseudo-potentials and lattice parameters as for light ice XI, thus neglecting these effects. The dynamical matrix was computed on a 5$\times$5$\times$4 Monkhorst-Pack grid by perturbation calculation and further Fourier interpolated for the thermal diffuse scattering (TDS) and IXS maps. Diffuse scattering intensity maps and (Q-E) IXS maps were computed in first order approximation with locally developed software, following the well established formalism \cite{bosak_zkri_2012,xu_zkri_2005}.
For the calculation of diffuse scattering maps the lattice parameters were rescaled to the experimental lattice parameters of ice Ih; for the calculation of x-ray patterns the O-H bond length was fixed to a value of 0.95\AA, as observed by X-rays, by translation along the O-O line. Individual Debye-Waller factors were calculated from the lattice dynamics data and averaged to produce the isotropic values used in the following calculations (see Appendix). The calculated Debye-Waller factors for oxygen are about 20\% smaller than experimentally observed \cite{kuhs_CUP_1986} at low temperatures but deviate by ca. 60\% at 220K. The deviation at high temperature may partly be explained by anharmonic contributions.

For the Monte Carlo simulations 128 fully disordered model clusters were generated starting from 64$\times$64$\times$64 cells (2$^{22}$ protons) ice XI crystal, producing a sufficient number of self-closed random walks. Cyclic boundary conditions were imposed; closed walks were preferred to prevent the creation of residual (charged) defects. The static disorder of oxygen atoms was taken into account by displacing the oxygen atoms along the H-O-H angle bisectrix by 0.05 \AA \hspace{1pt} in consistence with experimental and theoretical values \cite{kuhs_CUP_1986,kuo_jcp_2005}. Hydrogen sites are split to 32 sublattices 64$\times$64$\times$64, oxygen sites are split to 48 sublattices of the same size to accelerate the calculations applying a fast Fourier transform. This split sites approach allowed the reduction of complexity from $O(N^6)$ to $O(N^3log^3(N))$, where $N$ is the number of cells along the edge of modelling box.
The intensity of total scattering, normalised per formula unit, is calculated as
\begin{equation}
\label{eq:ice_DS}
I(\boldsymbol{Q}) \approx I_{TDS} (\boldsymbol{Q}) + I_{stat} (\boldsymbol{Q}),
\end{equation}
where 
\begin{equation}
\label{eq:ice_TDS}
I_{TDS} (\boldsymbol{Q}) = \frac{\hbar}{Z_{dyn}} \sum_{j=1}^{3n} \frac{1}{2 \omega_{j}(\boldsymbol{q})} coth \left( \frac{\hbar \omega_{j}(\boldsymbol{q})}{2 k T} \right) \left\vert \sum_{d=1}^n \frac{f_d(\boldsymbol{Q})}{\sqrt{M_d}} e^{-W_{d}(\boldsymbol{Q})+i\boldsymbol{Q} \cdot \boldsymbol{r}_d} (\boldsymbol{Q} \cdot \boldsymbol{\sigma}_{d}^{j}(\boldsymbol{q})) \right\vert^2
\end{equation}
is the TDS intensity \cite{xu_zkri_2005} and
\begin{equation}
\label{eq:ice_stat}
I_{stat} (\boldsymbol{Q}) = \frac{1}{Z_{stat}} \left\vert \sum_{d=1}^N  f_d(\boldsymbol{Q})  e^{-W_{d}(\boldsymbol{Q}) +i \boldsymbol{Q} \cdot \boldsymbol{r}_d}  \right\vert^2
\end{equation}
corresponds to the contribution of disorder. 
Here, for $n$ atoms per primitive unit cell, $\omega_j(\boldsymbol{q})$ - frequency of mode $j$ at reduced momentum transfer $\boldsymbol{q} = \boldsymbol{Q} - \boldsymbol{\tau}$, $\boldsymbol{\sigma}_{d}^{j}(\boldsymbol{q})$ - $d$-site projected component of the 3$n$-dimensional normalized eigenvector of the phonon mode $j$ defined in periodic notations $\sigma^j(\boldsymbol{q} + \boldsymbol{\tau}) = \sigma^j(\boldsymbol{q})$, where $\boldsymbol{\tau}$ is an arbitrary reciprocal lattice vector, $f_d(\boldsymbol{Q})$ - atomic scattering factor of atom $d$ with mass $M_d$ and Debye-Waller factor $W_d(\boldsymbol{Q})$ at the position $\boldsymbol{r}_d$, $k$ - Boltzmann's constant. Note that for neutron scattering $f_d(\boldsymbol{Q})$ is replaced by the $Q$-independent neutron scattering length $b_d$. $Z_{dyn}$ is the number of formula units in the primitive cell (4 for ice XI) and $Z_{stat}$ is the total number of formula units used in simulation (here $2^{28}$).

\section{Results and discussion}
\begin{figure}
\centering
\includegraphics[width=1.0\textwidth]{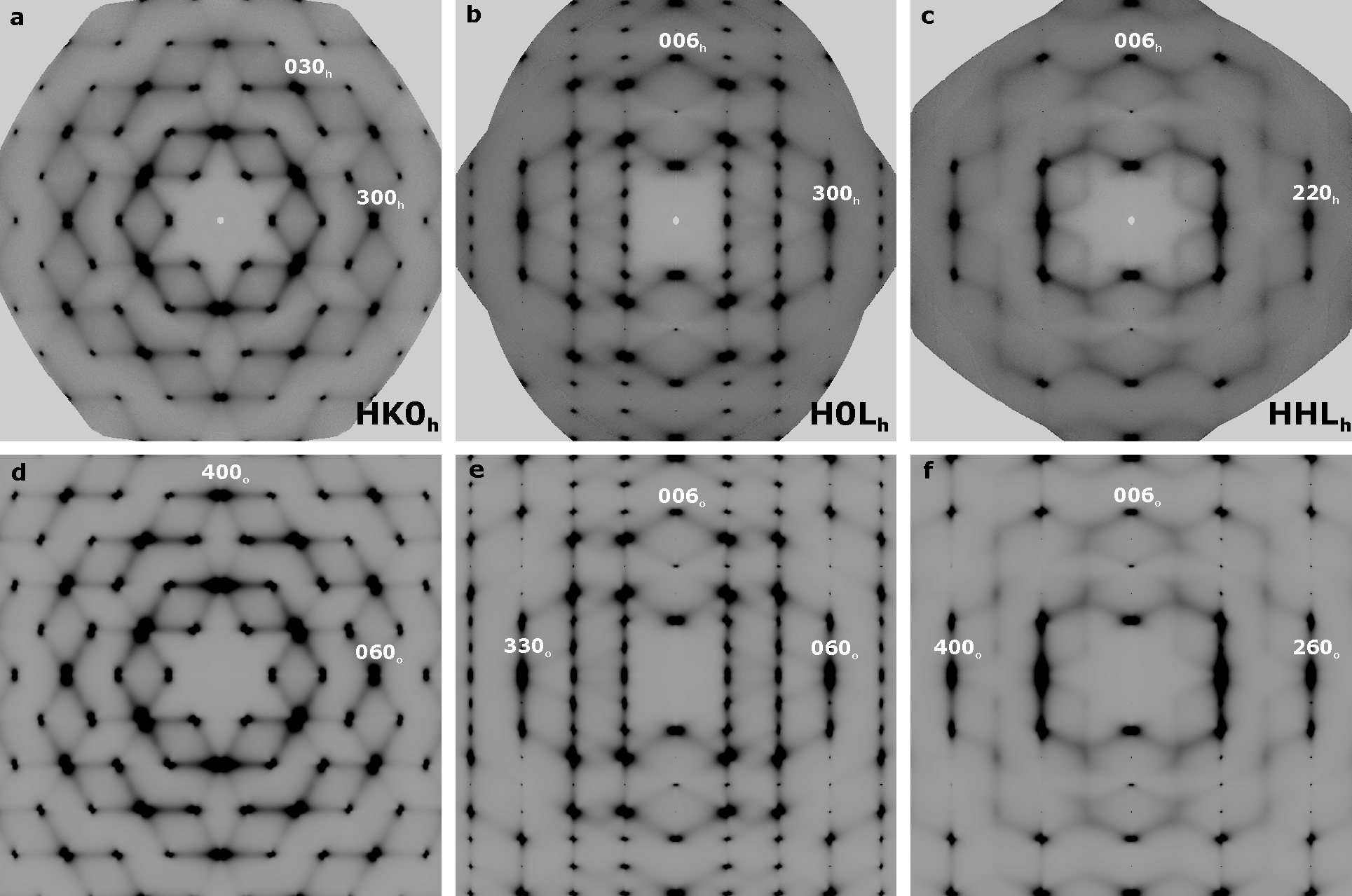}
\caption{\label{fig:ice_diffuse_maps}Diffuse x-ray scattering intensity maps of H$_2$O ice Ih at 175 K (a, b, c) compared to the \textit{ab initio} calculation for the ice XI approximant (d, e, f). For the cell relationships see Fig. \ref{fig:ice_IXS_maps}c. The HK0$_h$ plane in ice ih corresponds to the HK0$_o$ plane in ice XI. The H0L$_h$ plane corresponds to both HHL$_o$ (left half of panel e) and 0HL$_o$ (right half of panel e) planes and the HHL$_h$ plane to both H0L$_o$ (left half of panel f) and H 3K L$_o$ (right half of panel f) planes. Note the inequivalence of left and right halves of panels (e) and (f) and the absence of a six-fold axis in panel (d).} 
\end{figure}
Fig. \ref{fig:ice_diffuse_maps} shows representative experimental and theoretical diffuse scattering maps. The three high symmetry reciprocal space sections HK0$_h$, H0L$_h$, and HHL$_h$ of hexagonal ice Ih (labelled h) are compared to the corresponding reciprocal space sections of orthorhombic ice XI (labelled o). The relationships between the hexagonal and orthorhombic unit cell are illustrated in Fig. \ref{fig:ice_IXS_maps}c. The HK0$_h$ plane in ice ih corresponds to the HK0$_o$ plane in ice XI whereas the H0L$_h$ plane corresponds to both HHL$_o$ and 0HL$_o$ planes and the HHL$_h$ plane to both H0L$_o$ and H 3K L$_o$ planes. We note an excellent agreement between experiment and calculation: all features are well reproduced. A detailed inspection of the calculated TDS maps reveals some asymmetry in the scattering intensity distribution and the absence of the six fold symmetry in the HK0$_o$ plane. The inequivalence of the two orthorhombic planes corresponding to H0L$_h$ and HHL$_h$ of the hexagonal phase is clearly visible but does not affect the shape of the intense features. The intense features are very structured, in agreement with previous semi-quantitative data \citep{owston_ac_1949}. From this we can conclude that at least for low-energy modes the \textit{ab initio} calculation for the proton-ordered ice XI approximant reproduces quite well the dynamics of ice Ih both in terms of eigenvalues and oxygen eigenvectors since the TDS intensity in the low-energy part is $\sim\omega^{-2}$ (see Eq. \ref{eq:ice_TDS}) and therefore low-energy phonon modes provide the dominant contribution. 

\begin{figure}
\centering
\includegraphics[width=1.0\textwidth]{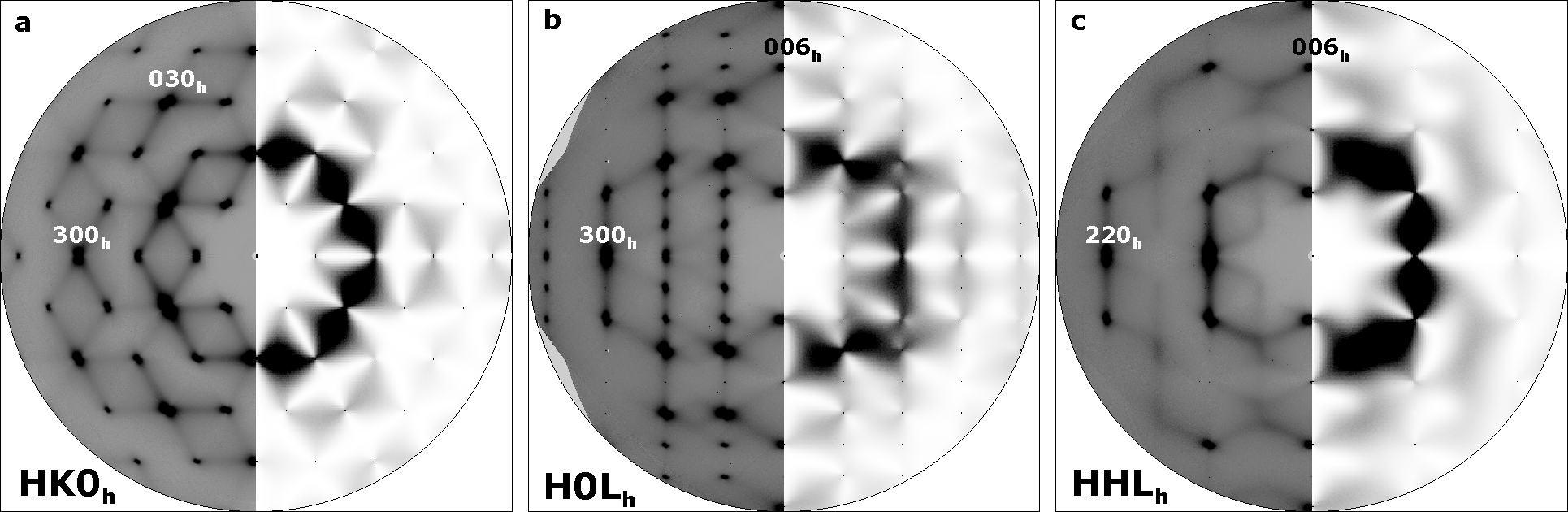}
\caption{\label{fig:ice_DS_static}Diffuse x-ray scattering intensity maps of H$_2$O ice at 175 K (left subpanels) compared to the calculated static x-ray contribution (right subpanels): (a) HK0$_h$ plane; (b) H0L$_h$ plane; (c) HHL$_h$ plane.} 
\end{figure}

\begin{figure}
\centering
\includegraphics[width=1.0\textwidth]{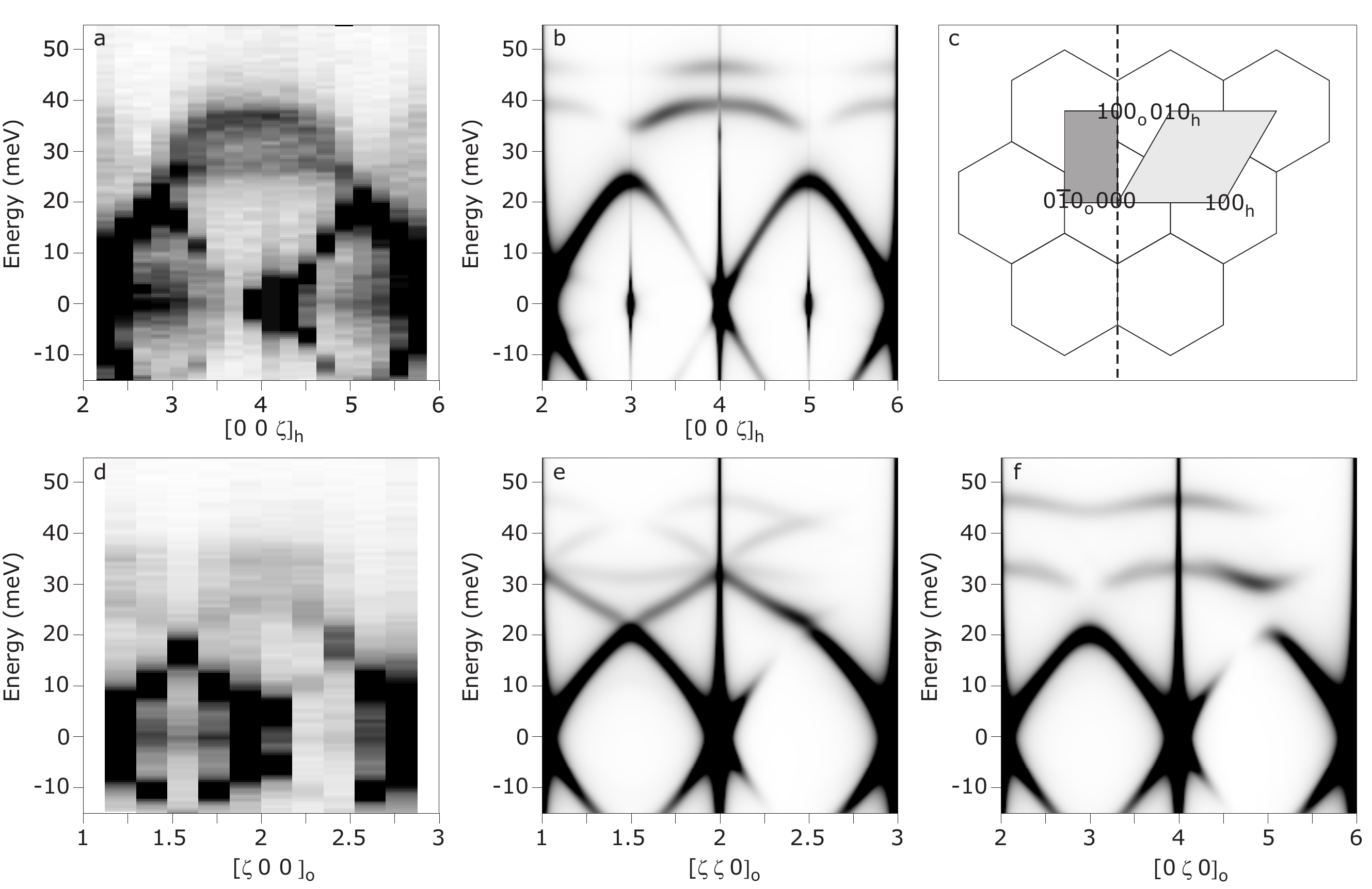}
\caption{\label{fig:ice_IXS_maps}(Q-E) IXS intensity maps of  H$_2$O ice Ih at 175 K for selected directions (a, d) compared to the \textit{ab initio} calculations (b, e, f). Both (e) and (f) correspond to (d) as those directions would become degenerate in disordered Ih phase. Panel (c) illustrates the relationship between the hexagonal (h) unit cell of ice Ih and the orthorhombic (o) unit cell of ice XI.} 
\end{figure}

The static disorder contribution to the diffuse scattering as computed from the Monte Carlo simulation is compared to experimental diffuse scattering in Fig. \ref{fig:ice_DS_static}. The intensity of the calculated patterns is magnified for best visualisation of the diffuse features. The absolute contribution of static disorder is very weak compared to TDS. We note that the regions with small contribution from static disorder roughly coincide with the minima of TDS. Consequently, there is no region where the static component could be observed separately.

Selected directions in reciprocal space were investigated by IXS and compared to the calculation in Fig. \ref{fig:ice_IXS_maps}. We note that the intensity distribution is well described by the calculation. Note that the [$\zeta$ $\zeta$ 0]$_o$ and [0 $\zeta$ 0]$_o$ directions in ice XI become degenerate in ice Ih. The experimental intensity map Fig. \ref{fig:ice_IXS_maps}d can be described by the average of the intensities of the two directions in ice XI, except for the 200$_h$ point. Close to the Bragg peaks the intensity contribution from the calculated acoustic phonons increases significantly and the convolution with the experimental resolution function results in very high intensity over a large energy range, seen as vertical lines in Fig. \ref{fig:ice_IXS_maps}b, e and f. The measured phonon energies are systematically lower than the calculated ones for ice XI. The presented IXS maps provide the only available data on experimental phonon dispersions of H$_2$O ice except for one branch published in \citep{ruocco_prb_1996} and serve as proof that the lattice dynamics in the low energy region of ice Ih can be described in good approximation by the lattice dynamics of the ordered phase. 

\begin{figure}
\centering
\includegraphics[width=0.8\textwidth]{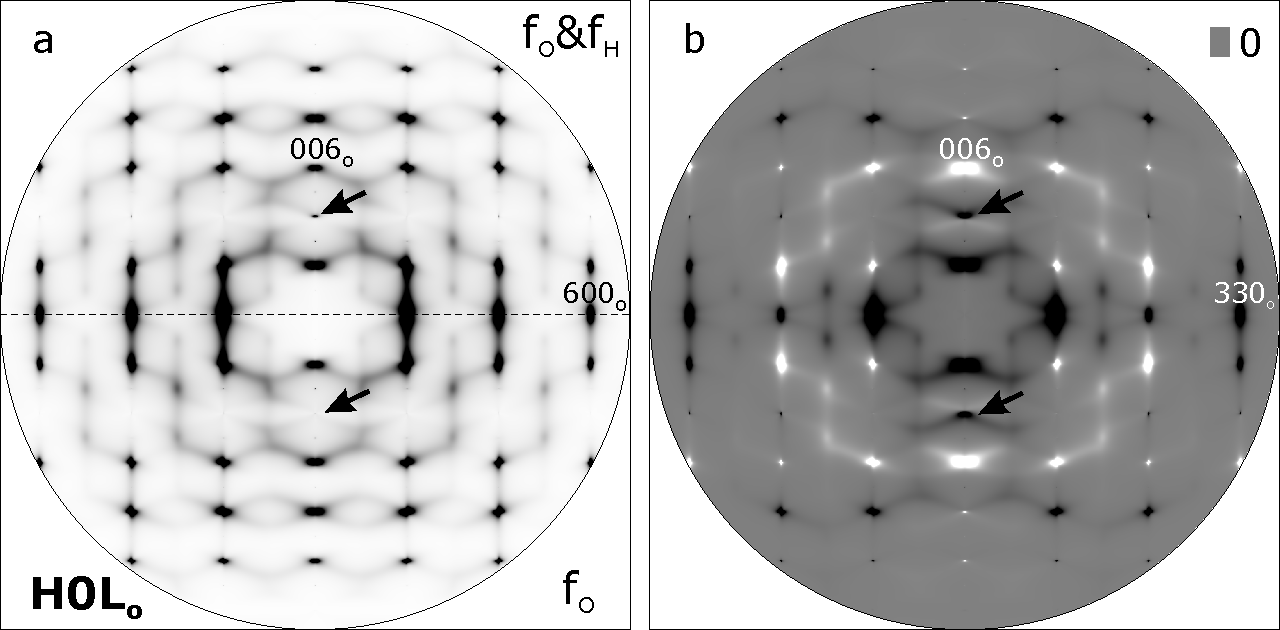}
\caption{\label{fig:ice_TDS_hydrogen}(a) Theoretical diffuse x-ray scattering intensity H0L$_o$ map of ice XI (equivalent to HHL$_h$ of ice Ih) at 175 K using both hydrogen and oxygen scattering factors (upper half) and oxygen scattering factor only (lower half). (b) Difference between the two maps of panel (a). The signal amplitude is multiplied by 4 and the zero difference is projected to medium grey (0x808080 color). To be compared with Fig. \ref{fig:ice_diffuse_maps}e. Arrows point towards the region of interest around the 004$_o$ node, an asymmetry can be noted in the difference map.} 
\end{figure}

The asymmetry of scattering intensity in the proximity of the 004$_h$ reflection and, to some extent in the proximity of the other reflections in the HK4$_h$ layer, is due to the hydrogen contribution to the scattering factor, as illustrated in Fig. \ref{fig:ice_TDS_hydrogen}. The TDS intensity distribution of the H0L$_o$ plane of ice XI (corresponds to HHL$_h$ in ice Ih) is calculated with and without the contribution of hydrogen scattering. The scattering around the 004$_o$ reflection disappears if the hydrogen contribution is neglected. Diffuse scattering around this reflection is indeed observed, see Fig. \ref{fig:ice_diffuse_maps}c. The observation that IXS and TDS are sensitive to the scattering from hydrogen is not fully unexpected, but can be considered as non-trivial experimental observation.

The calculated frequencies in the low-energy part of the spectrum are regularly higher than experimentally determined (Fig. \ref{fig:ice_IXS_maps}). This is in line with the available density of states measurements \citep{klug_prb_1991,schober_pb_1997}. Even within the above assumptions, our model provides valuable information, for example, for the estimation of purely phonon Debye-Waller factors and separation of the disorder contribution (so called "static Debye-Waller factor"). 

\begin{figure}
\centering
\includegraphics[width=1.0\textwidth]{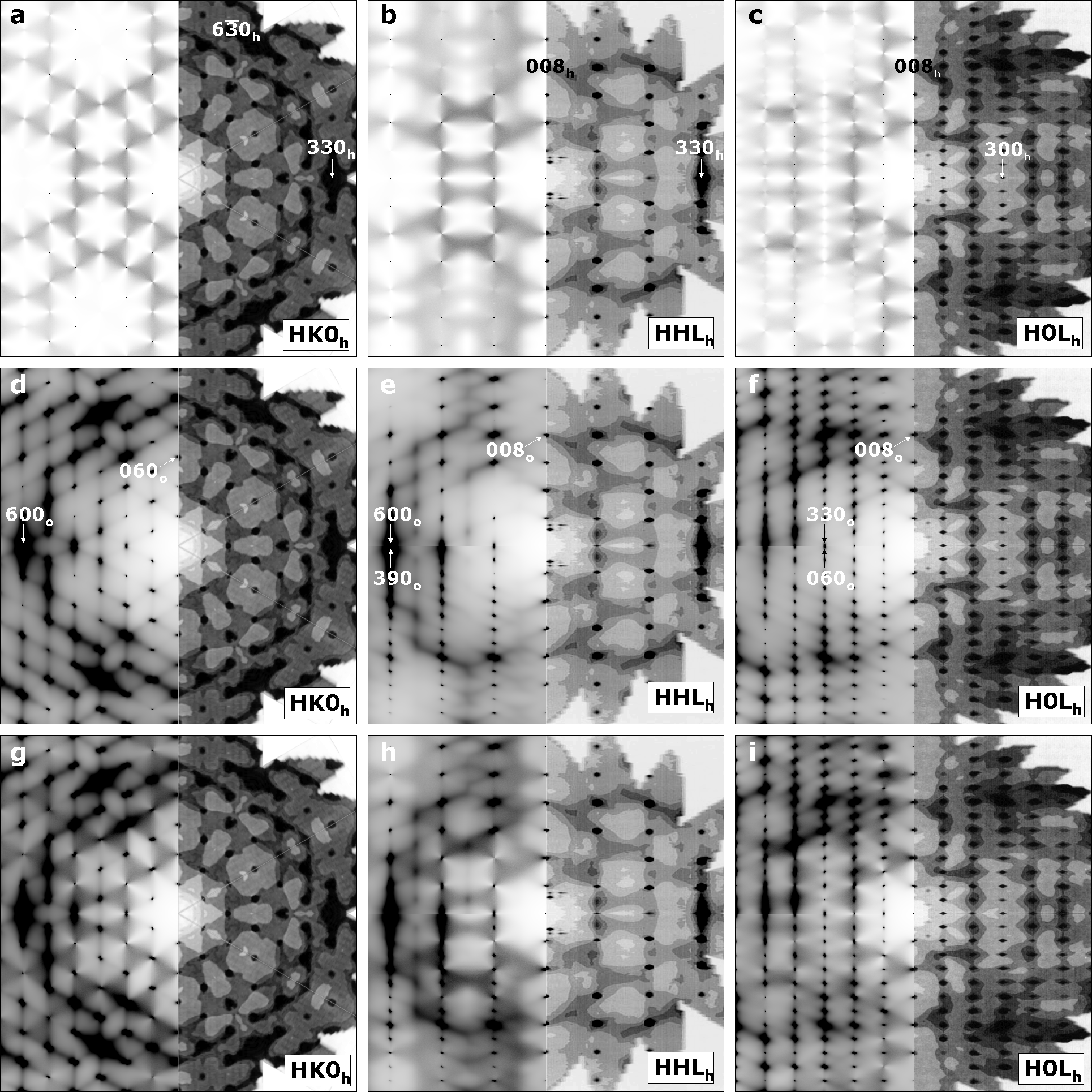}
\caption{\label{fig:ice_neutron}Diffuse neutron scattering intensity maps of H$_2$O ice Ih (right sub-panels, 10 K for HK0$_h$ plane and 20 K for H0L$_h$ and HHL$_h$ planes) \cite{li_pmb_1994} compared to the model calculations at 20 K (left sub-panels): including static displacements only (a, b, c), including TDS only (c, d, e), and including static displacement plus inelastic component (g, h, i). The same intensity scale is applied for all model plots.} 
\end{figure}

Our model was further used for the revision of published neutron scattering data \cite{beverley_pcb_1997}. Previously, the experimental patterns were modelled via a reverse Monte Carlo procedure, and the authors had to include additional frozen-in displacements to their model, while the phonon contribution was excluded based on the over interpretation of previously observed elastic scattering patterns \cite{schneider_jpc_1980}. Circumventing the natural limitations of Monte Carlo methods, we can now approximate the observed scattering as a sum of thermal diffuse scattering and disorder-related component, corresponding to Pauling's model. Fig. \ref{fig:ice_neutron} illustrates the results of such modelling, where the experiment \cite{li_pmb_1994} is compared to the thermal diffuse scattering maps, to the static contribution, and to the sum of above - taken for high-symmetry sections on the same scale.

\begin{figure}
\centering
\includegraphics[width=1.0\textwidth]{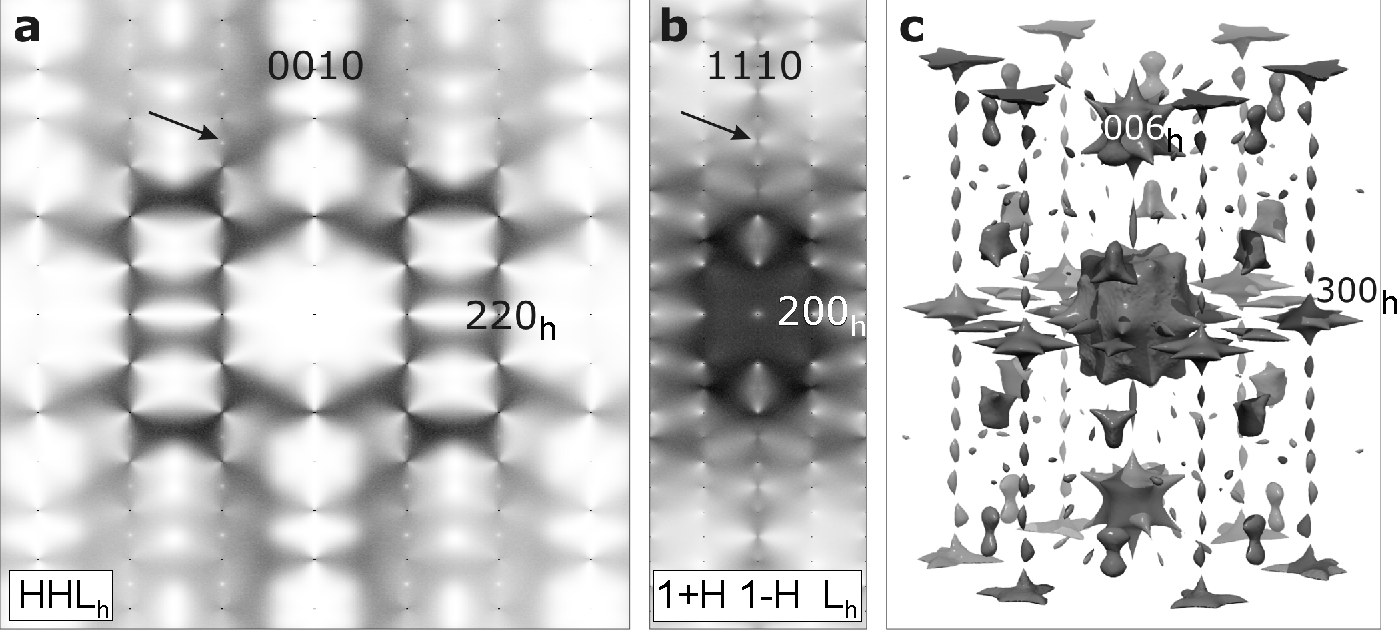}
\caption{\label{fig:ice_neutron_static}Calculated diffuse neutron scattering intensity maps for static displacements only. Arrows denote the low-intensity bow-tie-like feature, which can manifest as anti-node in the HHL$_h$ section. (a) HHL$_h$ cut; (b) 1+H 1-H L$_h$ cut; (c) isosurface representation of low-intensity features (I$_{max}$-I).
} 
\end{figure}

During the preliminary modelling of neutron scattering from static disorder, we noted the appearance of sharp local minima for forbidden Bragg node positions in some sections of reciprocal space (see Fig. 6a). Further analysis shows that these spots correspond to a bow-tie-like distribution of intensity in 3D, with the singularity point lying in the image plane (see Fig. \ref{fig:ice_neutron_static} representing two orthogonal sections in panels a and b). Thus, for any finite section thickness a local intensity minimum appears, while it would not be present for an infinitely sharp layer (Fig. \ref{fig:ice_neutron_static}a has to be compared to Fig. \ref{fig:ice_neutron}b, where the integration layer thickness is a factor 4 smaller). The singular points can be described by dipolar or power-law correlations in a non-divergent field obtained from a coarse graining of a local variable mapped to the ice rule degrees of freedom \cite{villain_ssc_1972}. Singularities of this kind were also observed in ice-rule ferroelectrics \cite{youngblood_prb_1981} and frustrated systems \cite{henley_arcmp_2010}. The disorder-related intensity distribution in 3D space is governed solely by Pauling's ice rules. For visualisation purposes we have chosen here the isosurface representation, where low-intensity regions are confined to something like a Swiss cheese with the pores connected by bow-tie-like singular points (Fig. \ref{fig:ice_neutron_static}c).

\section{Conclusions}
The studied ice sample from Lake Vostok accretion layer is of exceptional crystalline quality. This is witnessed by the absence of any traces of stacking faults. Thanks to this fact we could obtain unambiguous new data on the lattice dynamics of ice Ih and further analyse them in terms of \textit{ab initio} lattice dynamics calculations based on the ordered polymorph ice XI. 
The x-ray diffuse scattering is found to be almost entirely due to phonons (thermal diffuse scattering) and therefore may serve as a valuable source of information on the lattice dynamics. Our study reveals that the x-ray TDS is sensitive to thermal vibrations of hydrogen. At variance, neutron diffuse scattering originates (with comparable contributions at low temperature) from lattice dynamics contribution and the static hydrogen disorder. The latter is successfully quantified using Pauling's ice rules only.
The lattice dynamics of the disordered ice crystal can be compared to what one could expect for the ordered analogue. The phonon eigenvectors are similar for both ordered and disordered ice as follows from the experimental diffuse maps in reciprocal space for ice Ih compared with \textit{ab initio} calculations for ice XI. The experimental low-energy part of the phonon dispersion in high symmetry directions for ice Ih is regularly lower than calculated for ice XI.

\section*{Acknowledgement}
The authors would like to thank Keith Refson, Tom Fennell and Alessandro Mirone for fruitful discussions and comments on the manuscript.

\clearpage

\appendix
\setcounter{section}{1}
\section*{Appendix}

\begin{figure}
\centering
\includegraphics[width=0.6\textwidth]{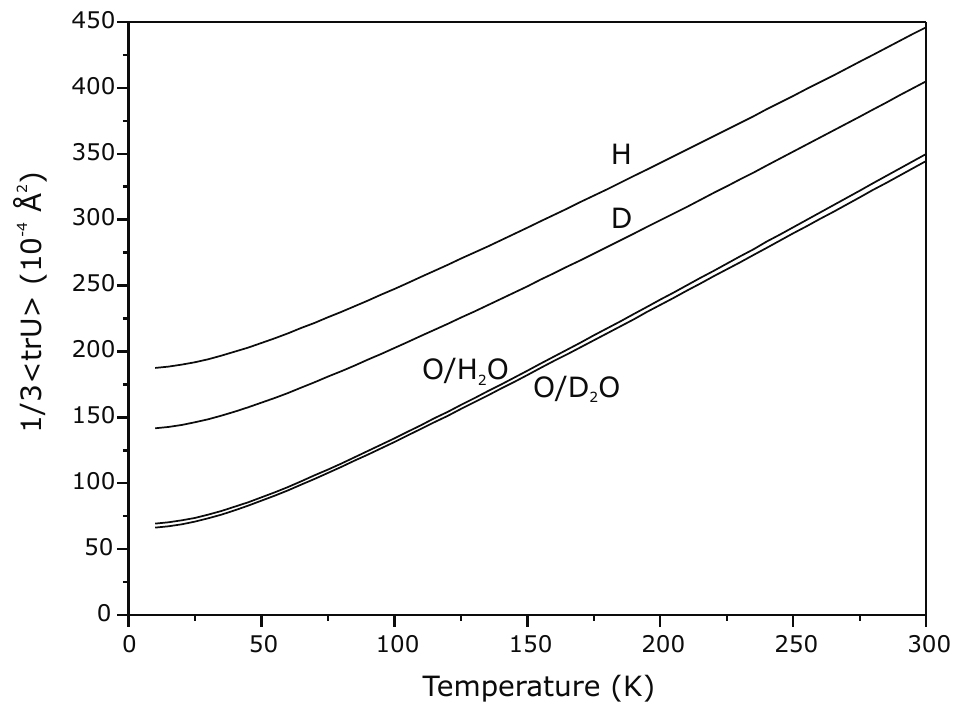}
\caption{\label{fig:ice_DW}Temperature dependence of the calculated Debye-Waller factors for the different species in H$_2$O and D$_2$O ice XI polymorphs.
} 
\end{figure}

\begin{figure}
\centering
\includegraphics[width=1.0\textwidth]{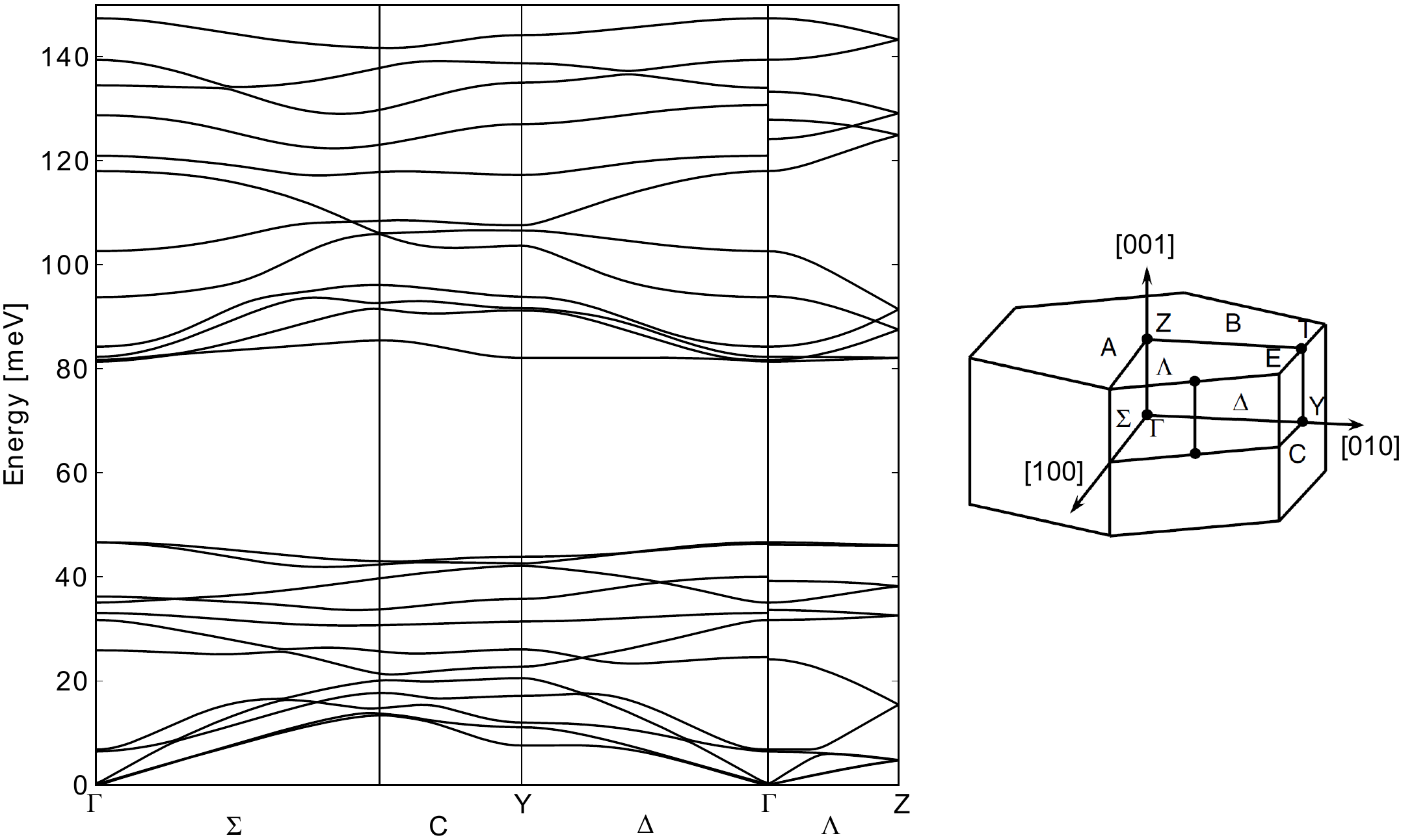}
\caption{\label{fig:ice_XI_dispersion}Calculated dispersion relations of ice XI along the indicated high symmetry directions.
} 
\end{figure}

Individual Debye-Waller factors were calculated from the dynamical matrix and averaged over all positions of the same atomic species in order to produce isotropic values. The temperature dependence of the averaged Debye-Waller factors is shown in Fig. \ref{fig:ice_DW}. The calculated dispersion relations of ice XI along some high symmetry directions are presented in Fig. \ref{fig:ice_XI_dispersion}.

\clearpage

\def\newblock{\hskip .11em plus .33em minus .07em}
\bibliographystyle{unsrturlabbrv}
\bibliography{references_ice}

\end{document}